# Experimental Test for Kinetic Quantum Gravity Theory

## Fran De Aquino

Maranhao State University, Physics Department, S.Luis/MA, Brazil.

We propose the following additional experiment to check the possibility of Gravity Control.

According to the Eq.(24) of *Kinetic Quantum Gravity* ( physics/ 0212033) the gravitational mass of a particle is changed when it absorbs or *emits* a photon. The Eq.(30), tell us that the gravitational mass of the electron can become *negative* if it emits a high-energy photon (γ-ray) with frequency $f > m_e c^2/h = 1.2 \times 10^{20} Hz$. There are several processes to make an electron emits a γ-ray. The proposed experiment is based on the *inverse* Compton effect. The Compton effect is well-known: a high-energy photon collides with an electron initially at rest, producing a photon with energy less than the energy of the incident photon and a recoil electron. In the *inverse* Compton effect a high-energy *electron* collides with a low-energy photon producing *a high-energy photon.* Thus consider the arrangement presented in Fig.1 where electrons are emitted from a 100 MeV *Betatron* with velocity =0.999986$c$ and collide with *infrared* photons (wavelength = 10.6μ$m$ ) inside a evacuated tube. As shown in Fig.1, φ+θ=30° where θ≅0° . Thus, the Compton effect theory predicts that after the collision each electron emits a photon with frequency $f$ given by:

$$f \cong p'_e(c/h) \cong \left( m_e v'_e / \sqrt{1 - (v'_e/c)^2} \right)(c/h) \cong$$
$$\cong 2.3 \times 10^{22} Hz$$

At this moment, the gravitational mass of the electron, $m_{ge}$ , according to the Eq.(30), becomes

$$|m_{ge}| \cong -\frac{2hf}{c^2} \cong -371.7 m_e$$

Therefore, according to the Eq.(48), the gravitational force upon the electron (electron-Earth) will be given by

$$\vec{F} = -G\frac{|M_\oplus||m_{ge}|}{r^2}\vec{\mu} = |m_{ge}|\left( -G\frac{|M_\oplus|}{r^2}\vec{\mu} \right) = |m_{ge}|(-g\vec{\mu}) =$$
$$= \left( -\frac{2hf}{c^2} \right)(-g\vec{\mu}) = \left( \frac{2hf}{c^2} \right)g\vec{\mu} = \left( \frac{2hf}{c^2} \right)\vec{g}_e$$

This means that the acceleration due to the gravity upon the electron becomes

$$\vec{g}_e = g\vec{\mu} \qquad (repulsion)$$

Before the collision: $\vec{g}_e = -g\vec{\mu}$ .

Thus, the electrons go up and strike the metal plate P with velocity $=\sqrt{2g_e y} \cong 4.4 m/s$ . Then a current of negative charge, $I_G$ , will be observed by the galvanometer G.

In order to the trajectory of the electrons be *vertical* , the diagram of the *inverse* Compton effect presented in Fig.1 must be *symmetric* of the diagram of Compton effect where a photon with frequency $f \cong 2.3 \times 10^{22} Hz$ strikes an *electron at rest* producing a photon with wavelength = 10.6μ$m$ (φ = 30° ) and giving to the electron a recoil velocity $v'_e \cong 0.999\ 986c$ with θ≅0° ( recoil angle).

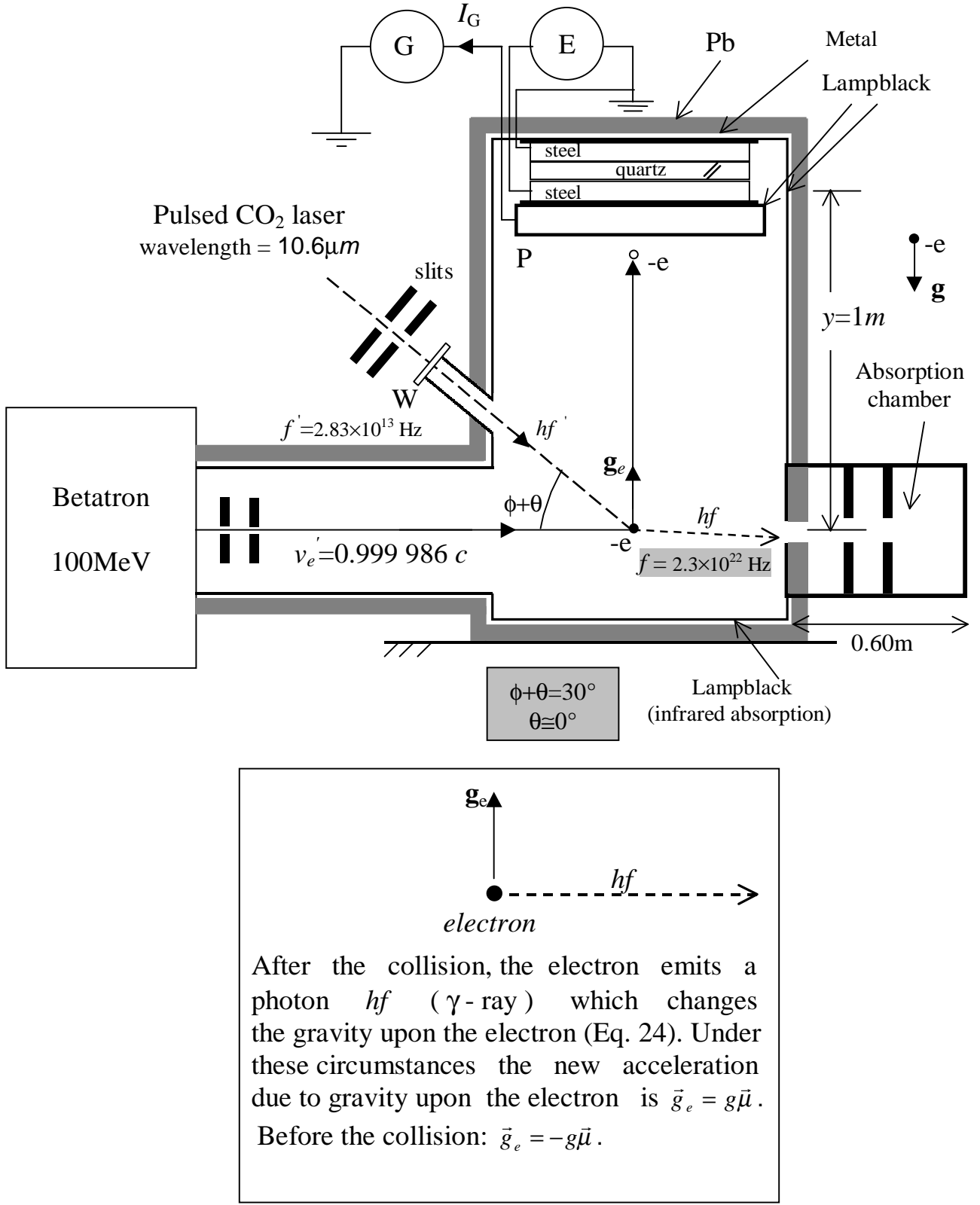

Fig.1 - Experimental arrangement for studying the gravitational behavior of electrons at the *Inverse* Compton Effect ( photon - electron collision ).

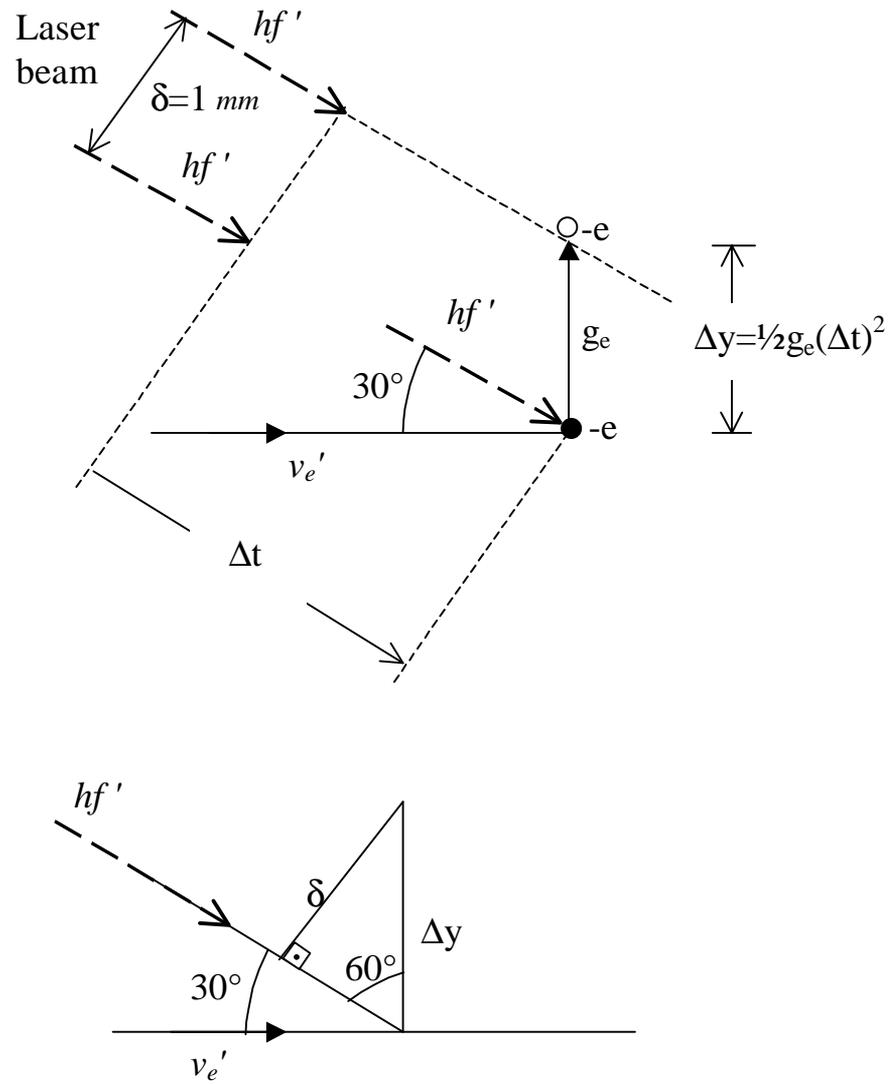

Fig.2 - If photons $hf'$ strike again with the electron of acceleration $\mathbf{g_e}$ it will be deviated of its vertical trajectory. So that this doesn't happen, the laser beam must be *pulsed* and in a such way that $\Delta t > \sqrt{2\Delta y/g_e} = \sqrt{2\ (\delta / \cos 30°\ )/g_e} \cong 0.8$ milliseconds ( $\delta = 1mm$ is the diameter of the laser beam. See figure above). Therefore, the interval among the pulses must be greater than 0.8 milliseconds . The $CO_2$ laser has a relatively long time nearly 1 milliseconds. This is sufficient for the present experiment.

The idea is that, *in the total absence of gravity* , the electron would remain *at rest* after the emission of the γ-ray, because this is its initial position of the electron at the Compton effect where the photon strikes the *electron at rest* (see Fig.3(a)).

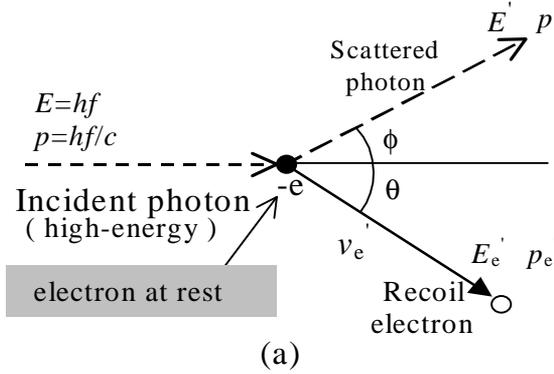

(a)

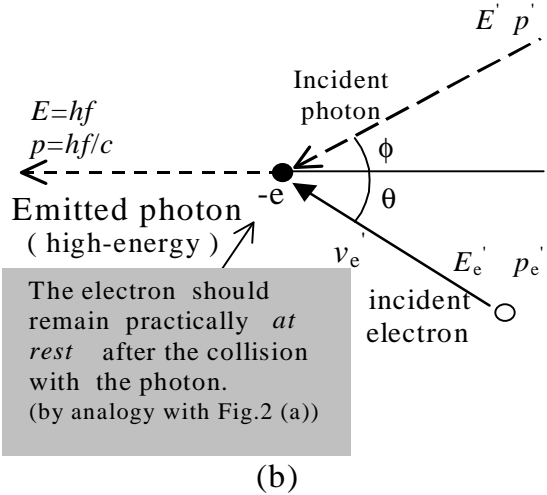

(b)

$$E^{'} = hf^{'}$$
$$p^{'} = hf^{'}/c$$
$$E_e^{'} = \sqrt{m_e^2 c^4 + p_e^{'2} c^2}$$
$$p_e^{'} = m_e v_e^{'} / \sqrt{1 - v_e^{'2}/c^2}$$

Fig. 3 - *Symmetric* Compton Effect.

Thus, at the present experiment, the only one motion of the electron after the collision would be caused exclusively by the gravity. Therefore, if the acceleration due to the gravity upon the electron has been inverted then the electrons go up right to the plate P.

At this experiment, in agreement with the Quantum Electrodynamics, we must consider the effects of the *vacuum polarization* because the energies of the electrons are very greater than their inertial energies at rest (at the direction of the betatron:

$$E_e = p_e^{'} c \cong m_e v_e^{'} c / \sqrt{1 - (v_e^{'}/c)^2} \cong 188.9 m_e c^2$$

at the direction of the plate P:

$$E_e = m_{ge} c^2 \cong (2hf/m_e c^2) m_e c^2 \cong 371.7 m_e c^2).$$

That is to say, we must consider the electrons inside their *polarization clouds* (electrons-positrons).

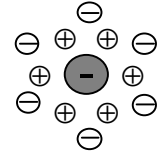

Fig.4 - Polarization of the *Vacuum*. Virtual *positrons* and virtual *electrons* around the electron.

Based on the Eq.(11), the well-known *Compton's equation* becomes

$$\lambda^{'} - \lambda = \frac{h}{|m_e| c}(1 - \cos \phi)$$

Note that now we have $|m_e|$ and not only $m_e$ as at the original expression. This means that the Compton effect is similar for *positrons* ( anti-matter).

The Eq.(30) tells us that the gravitational mass of the *positrons* and *electrons* are similar, and given by:

$$|m_{ge}| = -\frac{2hf}{c^2}$$

Thus, the gravitational forces upon the *positrons* will also be repulsive.

If $n_e$ is the number of electrons at the direction of the plate P, and around each one of these electrons there are $N_{ep}$ virtual positrons and virtual electrons then the total *momentum* $Q_p$ which they will transfer to the plate P will be given by:

$$Q_p = \left| M_{ge} \right| V = \left( n_e N_{ep} \right) m_{ge} \left| \sqrt{2 g_e y} \right. =$$

$$= \left( n_e N_{ep} \right) \left( \frac{2hf}{c^2} \right) \sqrt{2 g_e y}$$

and the force $\vec{F}_p$ upon the plate by:

$$\vec{F}_p = \left| M_{ge} \right| \vec{g} = \left( n_e N_{ep} \right) m_{ge} \left| \vec{g} \right. =$$

$$= -\left( n_e N_{ep} \right) \left( \frac{2hf}{c^2} \right) \vec{g} =$$

$$= \left( n_e N_{ep} \right) \left( \frac{2hf}{c^2} \right) \vec{g}_e$$

Then at the *electrometer E* will be observed a voltage proportional to that force.

We know that the *leptons* should have length scale less than $10^{-19} m$ [1]. This means that a electron has, at the maximum, "radius" $r_e \sim 10^{-19} m$. The plausible relation given by Brodsky and Drell [2] for the simplest composite theoretical model of the electrons, $\left| g - 2 \right| = r_e / \lambdabar_c$ or $\left| g - g_{DIRAC} \right| r_e / \lambdabar_c$, where $\lambdabar_c = 3.9 \times 10^{-13} m$ and $\left| g - 2 \right| = 1.1 \times 10^{-10} m$ [3] gives an electron radius $r_e \approx 10^{-22} m$. On the other hand, based on the *uncertainty principle,* we can evaluate the "radius" $\Delta r$ of the *polarization cloud* around the electron, i.e.,

$$\Delta r \sim \frac{\hbar c}{\Delta E} = \frac{\hbar c}{\left| m_{ge} \right| c^2} \cong \frac{\hbar}{371.7 m_e c} \cong 10^{-15} m$$

Mathematically, the particles maximum (electrons and positrons) inside the cloud is given by

$$\sim \left( \frac{\Delta r}{r_e} \right)^3$$

The quantity $N_{ep}$ should have the same order of magnitude ( due to the distribution of polarization), thus assuming that $r_e < 10^{-19} m$, we can write

$$N_{ep} \sim \left( \frac{\Delta r}{r_e} \right)^3 > 10^{12}$$

Therefore if $n_e \sim 10^{13}$; $(\sim 1 \mu A)$, the

values of $Q_p$ and $F_p$ will be the followings:

$$Q_p > 0.015 \, kg m / s \quad and \quad F_p > 0.033 \, N$$

This force is sufficiently intense to be detected by the electrometer $E$.

It is important to note that by increasing the intensity of the electrons beam from the betatron the force $F_p$ can be strongly increased.

Let us now consider a new situation for the arrangement presented in Fig.1. See Fig.5(a). We have introduced a cathode $C$ and an anode $A$ to accelerate (electrically) the electrons to the plate $P$ because now the direction of the plate $P$ is to 90° with respect to acceleration due to gravity $\vec{g}_e$.

Assuming that the acceleration due to the *electric field* between A,C is $\vec{a}_e \gg \vec{g}_e = -g\vec{\mu}$, the electrons strike the plate $P$ with velocity $= \sqrt{2 a_e y}$. The force $\vec{F}_p$ upon the plate is now

$$\vec{F}_p = \left| M_{ge} \right| \vec{a}_e = \left( n_e N_{ep} \right) m_{ge} \left| \vec{a}_e \right. =$$

$$= -\left( n_e N_{ep} \right) m_{ge} \left| \frac{e \vec{E}}{m_{ge}} \right| =$$

$$= -\left( n_e N_{ep} \right) e \vec{E}$$

It is easily verified that this system can works as a powerful *thrust engine* in *any direction*.

Note that the system presented in Fig.1 can also work as an injector of electrons and positrons with $\left| m_{ge} \right| = -2hf/c^2$ into a magnetic toroidal chamber where magnetic fields give to the electrons-positrons flux a *toroidal form* (analogous to the well-know system of confined toroidal plasma, *Tokamak*). See Fig.5(b).

The electrons-positrons toroidal flux will then have a *total gravitational mass* $M_{ge}$ such that

$$\left|M_{ge}\right| = -\left(n_e N_{ep}\right)\frac{2hf}{c^2}$$

Thus, assuming that the gravitational mass of the *toroidal chamber* is $M_{g(TOROID)}$ then if $\left|M_{g(TOROID)}\right| + \left|M_{ge}\right| < 0$ the system goes up ( with respect to the ground ). When $\left|M_{g(TOROID)}\right| + \left|M_{ge}\right| = 0$ the system stop in the air (floating). On the other hand, increasing the *negative gravitational mass* of the electrons-positrons flux (by increasing $n_e$ ), we can make $\left|M_{g(TOROID)}\right| + \left|M_{ge}\right|$ too small. For example, if the total gravitational mass of the system is

$$\left|M_{g(TOROID)}\right| + \left|M_{ge}\right| = 1kg$$

the system can acquire an enormous acceleration

$$\vec{a}_S = \vec{F}_S \Big/ \left|M_{g(TOROID)}\right| + \left|M_{ge}\right|$$

with a relative small force $\vec{F}_S$ .

Note that this is highly relevant for the aerospace engineering.

Consider the idea for a propulsion system, based on gravity control, presented in Figs.(6),(7),(8). It reveals a new concept of aerospatial propulsion.

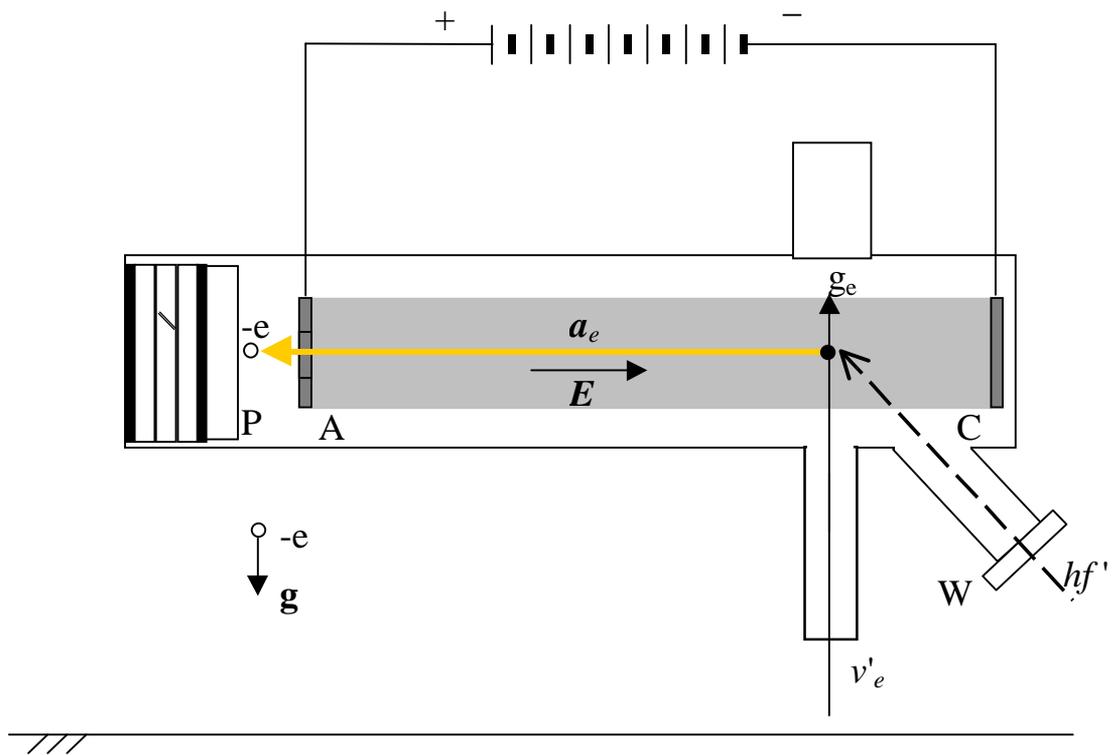

(a)

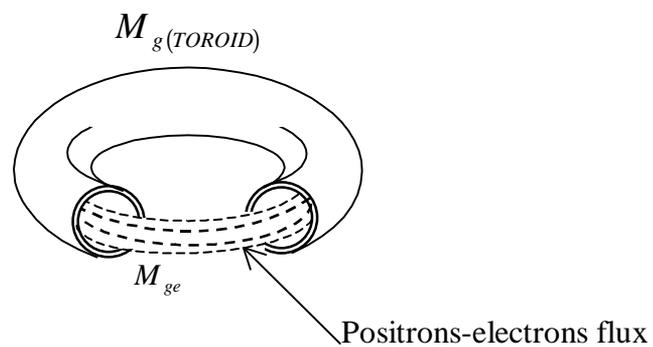

(b)

Fig. 5 - (a) A new situation for the arrangement presented in Fig.1. (b) Toroidal flux of positrons and electrons with *negative* gravitational mass.

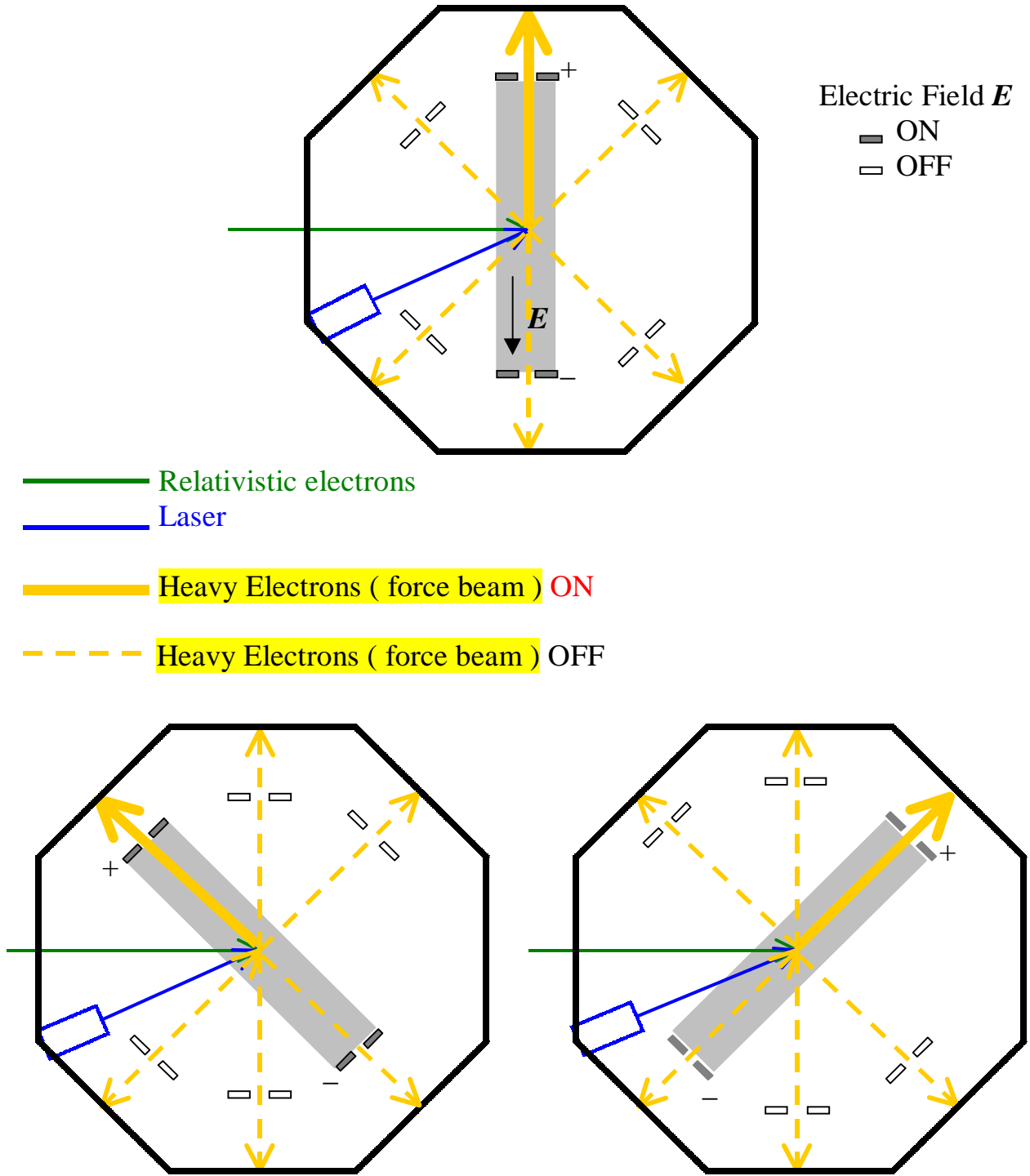

Fig. 6 - Propulsion in the direction of heavy electrons flow.

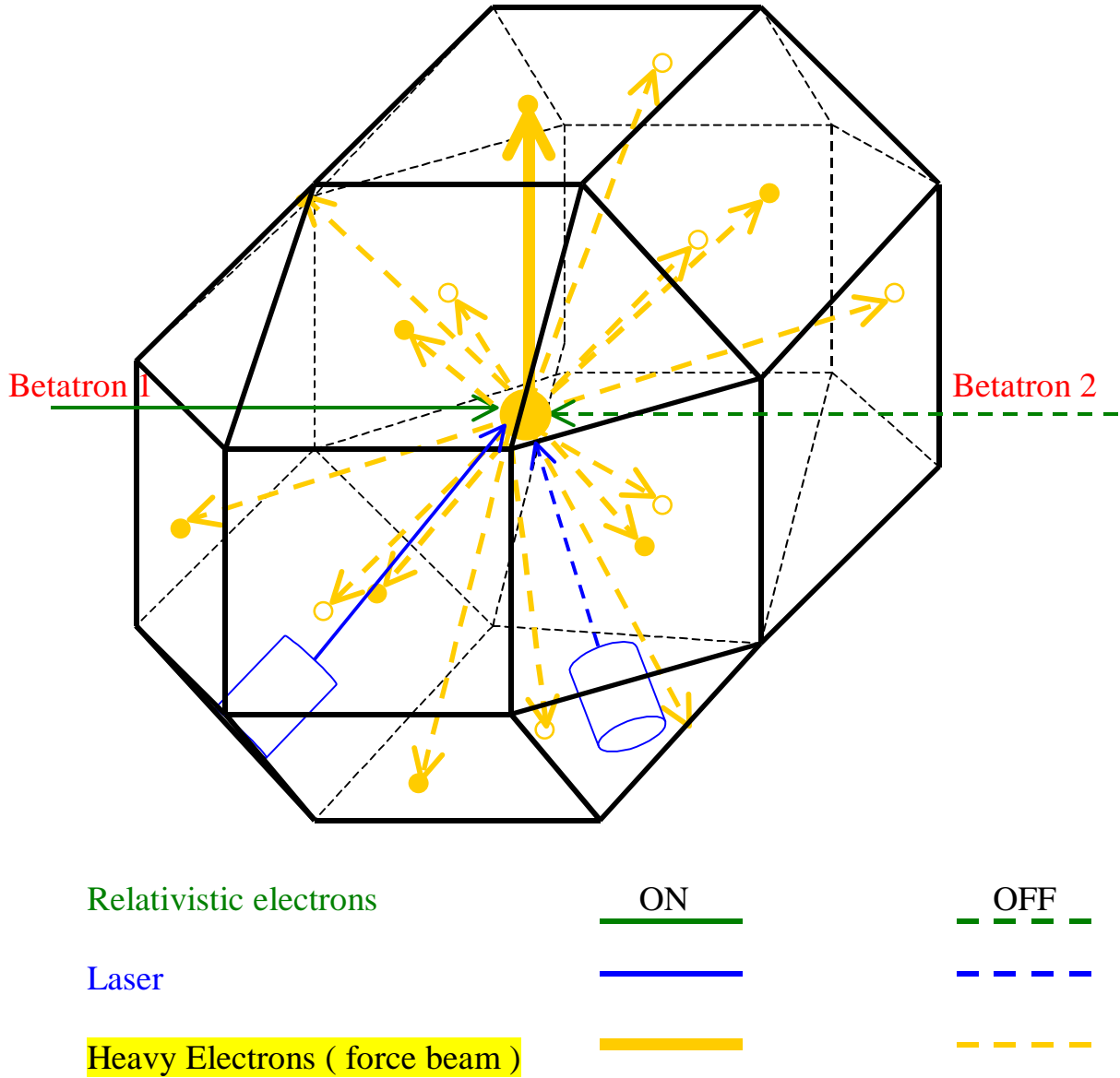

Betatron 1

Betatron 2

| Relativistic electrons | ON | OFF |
| --- | --- | --- |

Laser

Heavy Electrons ( force beam )

Fig. 7 - **The System Sun**

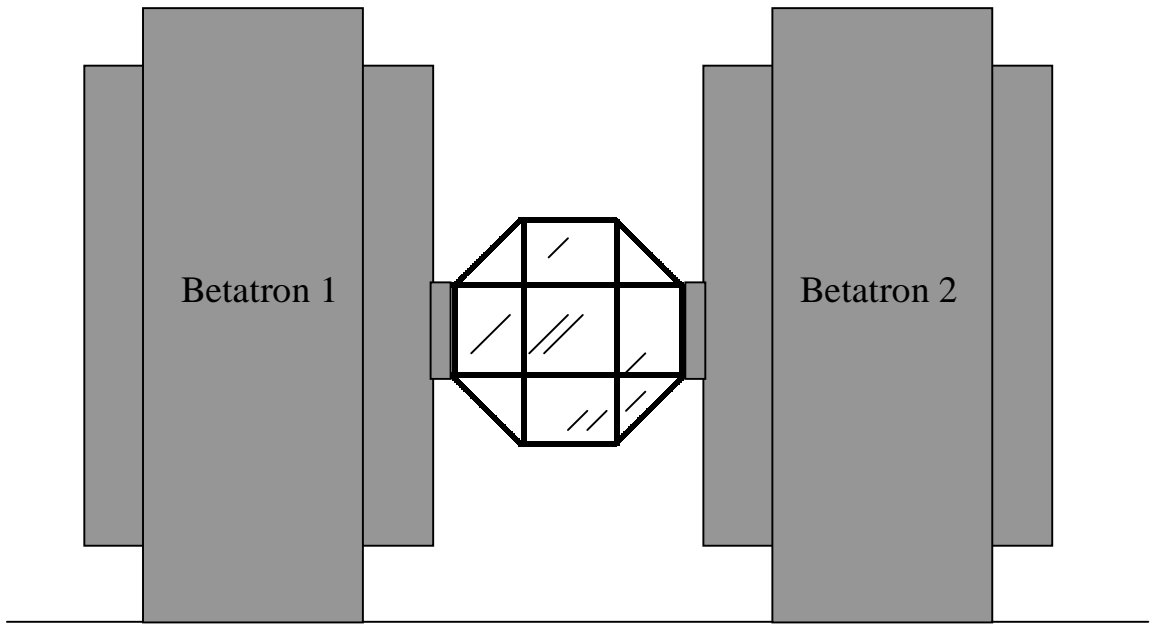

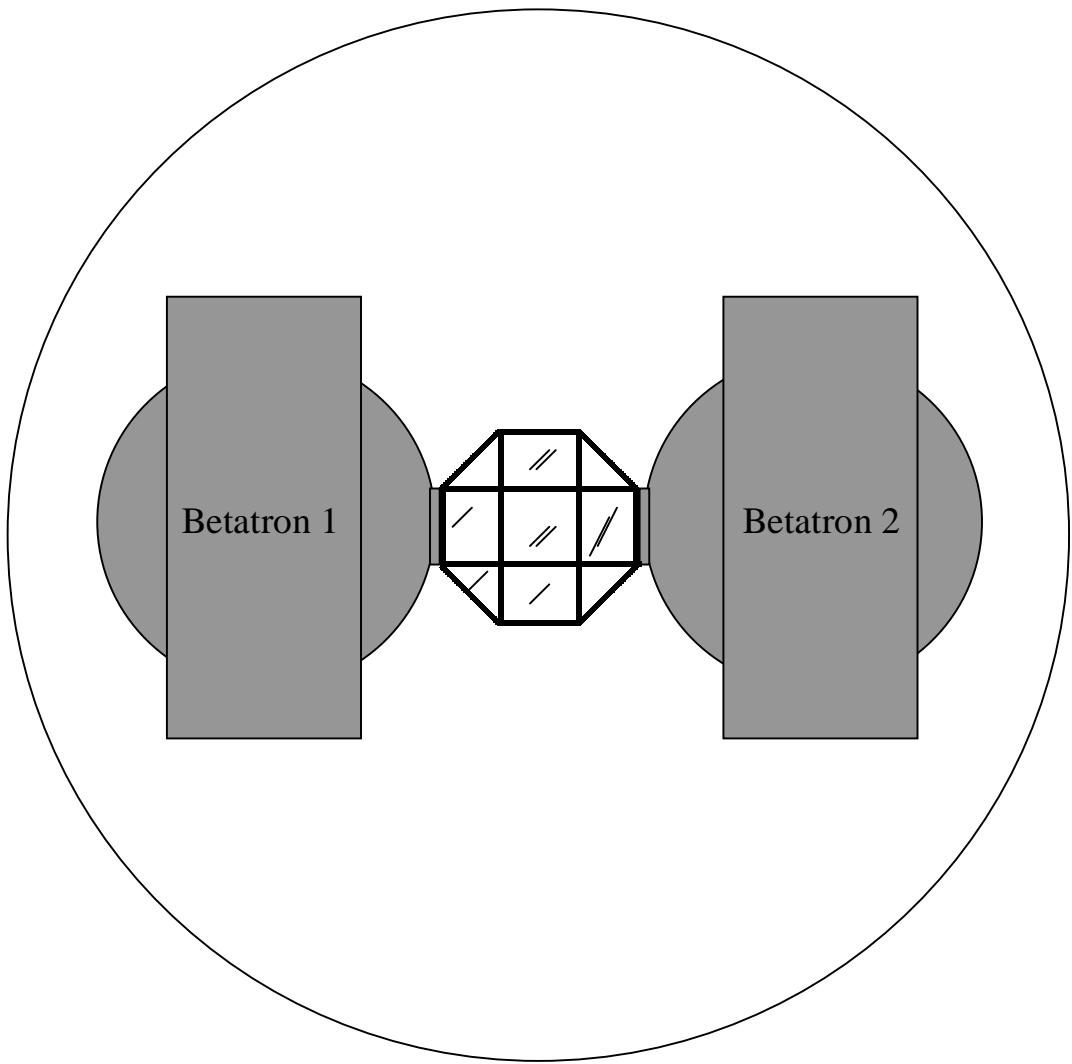

Fig.8 - Schematic Diagram of the System Sun.